\documentclass[twocolumn,superscriptaddress]{revtex4}
\usepackage{amssymb,amsmath}
\usepackage{graphicx}
\begin{document}

\title{Self-localization of laser induced tumor coagulation \\ limited by heat
diffusion through active tissue}
\author{B. Y. Datsko}
\author{V. V. Gafiychuk}
\affiliation{Institute for Applied Problems in Mechanics and Mathematics,
National Academy of Sciences of Ukraine}
\author{I. A. Lubashevsky}
\affiliation{Institute for Applied Problems in Mechanics and Mathematics,
National Academy of Sciences of Ukraine}
\affiliation{Theory Department,
General Physics Institute, Russian Academy of Sciences}
\author{A. V. Priezzhev}
\affiliation{Physics Department, Moscow State University}

\begin{abstract}
{} \vspace{\baselineskip}

We analyze the necrosis growth due to thermal coagulation induced by laser
light absorption and limited by heat diffusion into the surrounding live
tissue. The tissue is assumed to contain a certain tumor in the undamaged
tissue whereof the blood perfusion rate does not change during the action. By
contrast, the normal tissue responds strongly to increase in the tissue
temperature and the blood perfusion rate can grow by tenfold. We study in
detail the necrosis formation under conditions typical for a real course of
thermal therapy treatment, the duration of the action is taken about 5~minutes
when a necrosis domain of size about or above 1~cm is formed. In particular,
if the tumor size is sufficiently large, it exceeds 1~cm, and the tissue
response is not too delayed, the delay time does not exceed~1 min, then there
are conditions under which the relative volume of the damaged normal tissue is
small in comparison with the tumor volume after the tumor is coagulated
totally.
\end{abstract}

\maketitle

\section{Laser induced heat diffusion limited tissue
coagulation}

Thermal coagulation of living tissue caused by local heating due to laser
light absorption is one of the novel thermotherapy techniques of tumor
treatment which is currently undergoing clinical trials (see, e.g.,
\cite{I3}). Thermal coagulation is used to form a necrosis domain of desired
form for the removal of the malignant tissue. So mathematical modeling of the
necrosis growth is required, first, to find out the physical limitations and
the basic characteristics of the treatment and, second, to optimize the
therapy course. However, living tissue is extremely complex in structure,
thereby, for the adequate theoretical model to be developed and for the
mathematical modeling of the given process to be implemented reliably typical
limit cases should be singled out and studied individually.

\begin{figure}[h]
\begin{center}
\includegraphics[width=80mm]{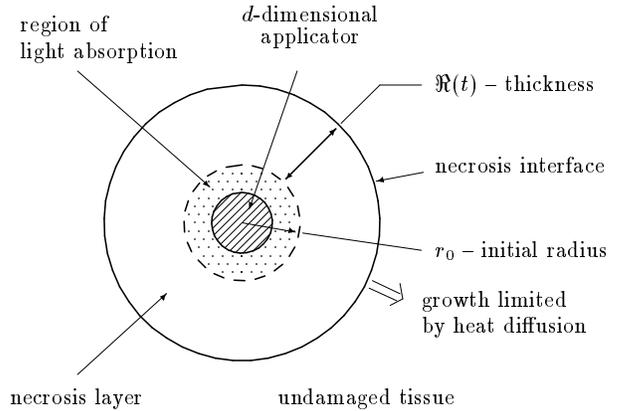}
\end{center}
\caption{The necrosis growth due to local thermal tissue coagulation limited
by heat diffusion in the surrounding undamaged tissue. $d=1$, 2 corresponds to
the surface plane applicator and the internal cylindrical one.} \label{ThC:F1}
\end{figure}

One of these cases is the laser induced heat diffusion limited tissue
coagulation~\cite{we9,we10}. It is characterized by the following features
(Fig.~\ref{ThC:F1}). Absorption of laser light delivered into a small internal
region of living tissue causes the temperature to attain such high values
(about or above $70~^{\circ }$C) that lead practically to immediate thermal
coagulation in this region. Heat diffusion into the surrounding live tissue
causes its further coagulation, giving rise to the growth of the necrosis
domain. In this case heat diffusion plays a significant role in the necrosis
growth because the necrosis size $\Re $ exceeds the depth of laser light
penetration into the tissue. Therefore, the temperature distribution
inevitably has to be substantially nonuniform and for the tissue to coagulate
at peripheral points heat diffusion should cause the temperature to grow at
these points. The latter property singles out the specific mode of thermal
coagulation under discussion from other possible types of thermotherapy
treatment and that is why we refer to the necrosis growth under the given
conditions as to thermal tissue coagulation limited by heat diffusion. In
particular, it turns out that the optimal implementation of this thermotherapy
mode is characterized by the formation of necrosis domains of size $\Re
\gtrsim 1$~cm and by the treatment duration of $t_{\text{course}}\sim \Re
^{2}/D\sim 2${--}$8$~min (where $D\approx 2\cdot 10^{-3}${\thinspace
cm}$^{2}$/sec is the tissue temperature diffusivity) \cite{we9,we10}.

Heat diffusion in the live tissue is affected substantially by blood
perfusion causing the heat sink \cite{CH80}. Thus, the characteristics of
the spatial distribution and the dynamics of the blood perfusion rate should
have a substantial effect on the necrosis growth limited by heat diffusion.
Therefore, in modeling this therapy mode one has to take into account the
tissue response to the temperature growth which can locally give rise to a
tenfold increase in the blood perfusion rate \cite{Song84}. The latter
effect, in particular, is responsible for a substantially nonuniform
distribution of the blood perfusion rate and visually manifests itself in a
red ring (``hyperemic ring'') appearing around the necrosis region during
the thermotherapy treatment.

In the previous papers \cite{we1,we2,we3,we4,we5,we6,we7,we8} basing on the
free boundary description we have developed a model for the heat diffusion
limited thermal tissue coagulation and studied the properties of the
corresponding necrosis growth. In particular, we have shown that the given
mode of thermal coagulation comprises two stages, fast and slow. At the former
the necrosis domain grows fast and its size attains values of order $\Re {\sim
}1$~cm. The latter is characterized by a substantially slower growth of the
necrosis domain. Exactly at this stage the necrosis growth is governed by the
competition between the heat diffusion into the surrounding undamaged tissue
and the heat dissipation caused by blood perfusion. Therefore, a sufficiently
strong tissue response to temperature variations which, in turn, gives rise to
an essential local increase in the blood perfusion rate has to affect the
necrosis formation. Namely, the higher is the local blood perfusion rate, the
smaller is the size of the necrosis domain.

Inside tumors the blood vessels are abnormal, in particular, they have lost
the ability to expand substantially in the response to tissue heating. As a
result, inside a tumor the blood perfusion rate can increase no more then
twice under strong local heating, although in the unaffected tissue the blood
perfusion rate inside a tumor typically exceeds the perfusion rate in the
normal tissue~\cite{Song84}. So, if the normal tissue exhibits the strong
response to temperature variations then the boundary of the necrosis domain
will penetrate deeper insider the tumor than in the normal tissue during the
thermal therapy course. Therefore in this case, first, the relative volume of
the damaged normal tissue can be less than the volume of the tumor after its
coagulation. Second, when the necrosis boundary reaches the tumor boundary the
necrosis growth should slow down. At first glance it would seem that the
necrosis growth is mainly confined to the tumor space. Exactly this phenomenon
is the subject of the present paper and will be called the self-localization of
the necrosis growth in active living tissue with a tumor.

\section{Free boundary model of the necrosis growth}

We study the necrosis formation caused by laser light absorption in living
tissue with a tumor applying to the following model \cite
{we9,we1,we2,we3,we4,we5,we6,we7,we8}. The laser light absorption causes heat
generation and, as a result, thermal coagulation of the tissue. For simplicity
sake the heat generation rate $q(\mathbf{r})$ is considered to be a beforehand
given function of the spatial coordinate $\mathbf{r}$ that is independent of
the tissue state and the time $t$ (certainly, for $t>0$ and $t=0$ is the
beginning of the action on the tissue). Besides, we assume the heat generation
rate $q(\mathbf{r})$ to be localized in a small region adjacent to the
applicator whose thickness is substantially less than the characteristic size
$\Re \gtrsim 1$~cm of the necrosis domain formed during a typical course of
thermotherapy treatment.

In modelling heat propagation in the tissue we single out three regions: the
necrosis domain $Q_{\text{n}}$, where the blood perfusion rate is equal to zero
\begin{equation}
j(\mathbf{r},t)=0\quad \text{for}\quad \mathbf{r}\in Q_{\text{n}}\,,
\label{ThC:e3.1}
\end{equation}
the undamaged tumor tissue $Q_{\text{t}}$, and the undamaged (live) normal
tissue $Q_{\text{l}}$. Inside the necrosis domain $Q_{\text{n}}$ the tissue
temperature $T(\mathbf{r},t)$ obeys the heat diffusion equation for solids:
\begin{equation}
c\rho \frac{\partial T}{\partial t}=\kappa \nabla ^{2}T+q,\quad \mathbf{r} \in
Q_{\text{n}}\,,  \label{ThC:e3.4}
\end{equation}
where $c$ and $\rho$ are the specific heat and density of the tissue, and
$\kappa $ is the cellular tissue heat conductivity. In the undamaged tissue
(including also the undamaged part of the tumor) the temperature is governed
by the bioheat equation (see also, \cite{BOOK}):
\begin{equation}
c\rho \frac{\partial T}{\partial t}=F\kappa \nabla ^{2}T-fc\rho
j_{v}(T-T_{a})+q,\quad \mathbf{r}\in Q_{\text{l}}\bigcup Q_{\text{t}}\,.
\label{ThC:e3.5}
\end{equation}
Here $T_{a}$ is the temperature of arterial blood in systemic circulation,
$j_{v}$ is the value of the blood perfusion rate $j$ locally averaged over
spatial scales
\begin{equation}
\ell _{v}(\mathbf{r})\sim \sqrt{\frac{\kappa }{c\rho f\,j_{v}(\mathbf{r})
L_{n}}}\sim 4\,\text{mm},  \label{ThC:e1.4}
\end{equation}
where $L_{n}=\ln (l/a)\approx 3-4$ and $l/a$ is the mean ratio of the
individual length to radius of blood vessels forming peripheral circulation,
the cofactor $f\sim 1/\sqrt{L_{n}}\sim 0.5$ takes into account the
counter-current effect \cite{WJ85,WXZE97}, and the factor $F=F(L_{n})\sim 2$
allows for the renormalization of the heat conductivity due to blood flow.
It should be pointed out that the scale $\ell _{v}(\mathbf{r})$ is exactly
the minimal scale on which equation~(\ref{ThC:e3.5}) is justified \cite{BOOK}
and the characteristic size of the necrosis domain formed during this mode
of thermal therapy can be estimated as $\Re \gtrsim \sqrt{L_{n}}\ell _{v}$:
\begin{equation}
\Re \gtrsim \sqrt{\frac{\kappa }{c\rho f\,j}}\,.  \label{exp:2}
\end{equation}

Finding the relationship between the averaged and true blood perfusion
rates, $j_{v}$ and $j$, we have to take into account that the scale $\ell
_{v}$ of averaging in its turn depends on the local value of $j_{v}$
(expression~(\ref{ThC:e1.4})). This dependence enables us to specify this
relationship in the form \cite{BOOK}:
\begin{equation}
j_{v}-\frac{\lambda _{v}\kappa }{c\rho }\nabla ^{2}\ln j_{v}=j,\quad
\mathbf{r}\in Q_{\text{l}}\bigcup Q_{\text{t}},  \label{ThC:e3.9}
\end{equation}
where $\lambda _{v}\sim 1/\sqrt{L_{n}}$ is also a constant of order unity.
Equation~(\ref{ThC:e3.9}) should be subjected to a certain boundary condition
at the interface $\Gamma $ of the necrosis domain because it makes no sense to
average the blood perfusion rate over the necrosis domain impermeable to
blood. The physical layer separating the necrosis domain and the undamaged
tissue where the local vascular network is not damaged is complex in structure
and contains a spatial increase of the blood perfusion rate from zero to the
value in the undamaged tissue. In order to avoid the problem of analyzing the
blood perfusion rate in this layer we take into account the following
simplifying circumstance. On one hand, the typical size of the necrosis domain
formed during a thermal therapy course and the characteristic length of
temperature variations are of the same order about 1~cm. So, particular
details of spatial variations in the blood perfusion rate on scale much less
than 1~cm are not the factor. On the other hand, the damaged part of the
vascular network located inside the necrosis domain is most probable to be
made up of an artery and vein having supplied previously this region with
blood as a whole and of shorter vessels formed by their branching. Therefore,
the region containing the vascular network part in which blood flow is
remarkably disturbed because of the necrosis formation does not exceed
substantially the necrosis domain. The latter enables us not to make
difference between the given layer and the interface $\Gamma $ and to deal
with a sharp jump of the blood perfusion rate at the necrosis interface. The
desired boundary condition imposed on the averaged blood perfusion rate
$j_{v}$ meets the requirement that the normal gradient of the averaged blood
perfusion rate be equal to zero at the interface $\Gamma $:
\begin{equation}
\left. \nabla _{n}j_{v}\right| _{\Gamma }=0\,.  \label{ThC:e3.11}
\end{equation}
We note that the adopted boundary condition will not hold if a large vessel
goes through the necrosis domain. However, the probability of this event is
small enough and this case should be analyzed individually.

Now let us specify the tissue response to temperature variations. Only the
undamaged normal tissue is considered to respond to local heating, inside the
undamaged tissue of the tumor the blood perfusion rate maintains the initial
value $j_{0}$ assumed for simplicity sake to be the same in the normal tissue
and the tumor
\begin{equation}
j(\mathbf{r},t)=j_{0}\quad \text{for}\quad \mathbf{r}\in Q_{\text{t}}\,.
\label{exp:1}
\end{equation}
Blood vessels can expand only to a certain extent as the temperature grows. So
when it becomes high enough, $T>T_{\text{vr}}$, the blood perfusion rate
$j(\mathbf{r},t)$ attains a large but finite value $j_{\text{max}}$ and remains
approximately constant. Keeping in mind this feature we describe the normal
tissue response to local heating by the equation:
\begin{equation}
\tau \frac{\partial j}{\partial t}+j\Phi (T)=j_{0}\quad \text{for}\quad
\mathbf{r}\in Q_{\text{l}}\,.  \label{ThC:e3.2}
\end{equation}
Here $\tau $ is the delay time of the tissue response and the function $\Phi
(T)$ is of the form
\begin{equation}
\Phi (T)=\left\{
\begin{array}{ll}
\alpha +(1-\alpha ){\dfrac{T_{\text{vr}}-T}{T_{\text{vr}}-T_{a}}} & \quad
\text{for}\quad T<T_{\text{vr}} \\
\alpha  & \quad \text{for}\quad T>T_{\text{vr}}
\end{array}
\right.   \label{ThC:e3.3}
\end{equation}
where $\alpha =j_{0}/j_{\text{max}}$ and the temperature at which the blood
vessels exhaust their ability to expand is estimated as $T_{\text{vr}}\approx
45$--$46^{\circ }$C.

Equations (\ref{ThC:e3.4}) and (\ref{ThC:e3.5}) governing the evolution of
the tissue temperature should be completed by the boundary conditions
relation the temperature field and the necrosis interface $\Gamma $. To
solve this problem we have developed the free boundary model that assumes
the temperature and the heat flux to have no jumps, i.e. the temperature
distribution meets the following boundary conditions
\begin{equation}
\left. T\right| _{\Gamma +0}=\left. T\right| _{\Gamma -0}\overset{\text{def}%
}{=}T_{\text{cg}}\,,\qquad \left. F\nabla _{n}T\right| _{\Gamma +0}=\left.
\nabla _{n}T\right| _{\Gamma -0}\,.  \label{ThC:e3.6}
\end{equation}
and the normal velocity of the interface $\Gamma $ is given by the
expression
\begin{equation}
\vartheta _{n}=\frac{\Im _{0}\omega _{0}\Delta }{\left| \nabla _{n}T\right|
_{\Gamma -0}}\exp \left[ \frac{T_{\text{cg}}-T_{0}}{\Delta }\right] \,.
\label{ThC:e3.7}
\end{equation}
Here the constant $\Im _{0}\sim 1$ and the function
\begin{equation}
\omega (T)=\omega _{0}\exp \left[ \frac{T-T_{0}}{\Delta }\right] \,.
\label{ThC:e1.3}
\end{equation}
is actually a convenient approximation of the Arrhenius dependence of the
thermal coagulation rate $\omega (T)\propto \exp \{-\frac{E}{T}\}$. The
available experimental data \cite{J94} for the temperature dependence of the
exposure time enable us to estimate the value of $\Delta $ as $\Delta \sim
3-5\,^{\circ }$C ($\Delta \simeq 3.6\,^{\circ }$C for pig liver at $%
T_{0}=60\,^{\circ }$C). Below in numerical calculations the dependence~(\ref
{ThC:e1.3}) will be taken in the form
\begin{equation}
\omega (T)=0.2\exp \left[ \frac{T-60}{3.6}\right] ~\text{min}^{-1}\,,
\label{ThC:e1.3a}
\end{equation}
where the temperature $T$ is in degrees Celsius.

\section{Results of numerical simulation. Self-localization of the necrosis
growth in living tissue with a tumor}

\begin{figure}[t]
\begin{center}
\includegraphics[width=70mm]{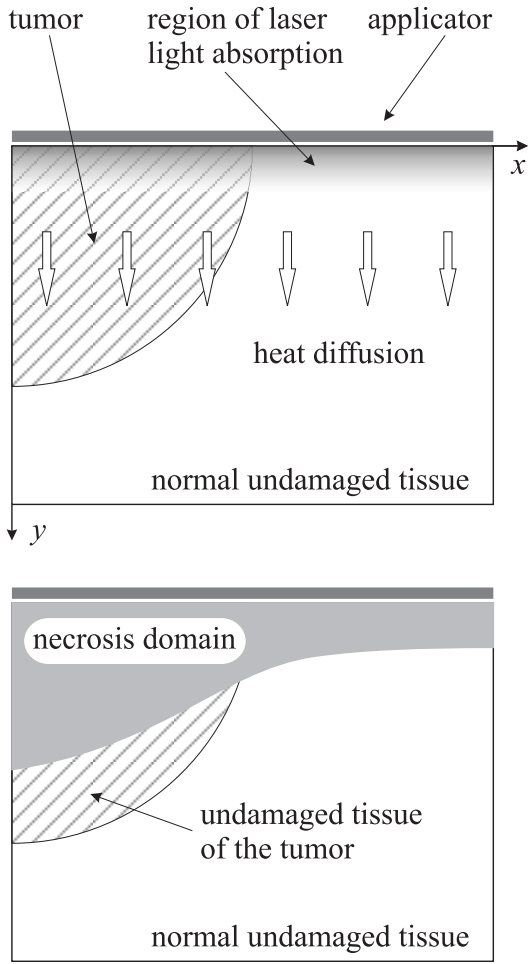}
\end{center}
\caption{The system where the necrosis formation has been studied numerically
based on the free boundary model. The $xy$-region of the plane and cylindrical
symmetry imitates the effect of the surface applicator and that of the
internal cylindrical one.}\label{Fig.2}
\end{figure}

The stated model has been used to analyze numerically the necrosis growth in
the case shown in Fig.~\ref{Fig.2}. The laser light absorption near the
applicator gives rise to the heat generation rate $q(y)$ uniform along the $x
$-axis and decreasing with $y$ as $q(y)\propto \exp (-y/l_{\text{ad}})$, where
the characteristic thickness $l_{\text{ad}}$ of the laser light absorption
layer is chosen less then the characteristic size $\Re \gtrsim
1$\thinspace cm of the necrosis domain formed during the thermal therapy mode
under consideration, $l_{\text{ad}} < \Re $. The intensity of laser light has
taken such values that the tissue temperature directly near the applicator,
$y=0$, be about 80 to 90 $^{\circ }$C ( $q(y)=2 \exp (-y)$ for large tumor and $q(y)=4 \exp (-4y)$ for small tumor). We have studied two opposite cases
specified by the relation of the scale $\Re $ (see expression~(\ref{exp:2}))
and the tumor size $\mathcal{L}$. For a small tumor, $\mathcal{L}<\Re $, the
self-localization phenomenon cannot come into being and this case is presented
for illustration only. For a large tumor, $\mathcal{L}>\Re $, there are
conditions under which the self-localization is pronounced as is demonstrated
directly in the present section. Besides, we have considered the $xy$-region
of plane geometry as well as cylindrical one in order to compare the necrosis
growth for applicators of different geometry, the surface plane applicator and
internal cylindrical one. In the latter case the radius $r_{0}$ of the
applicator also was assumed to be small in comparison with the scale $\Re $.

\begin{figure}[hf]
\begin{center}
\includegraphics[width=60mm]{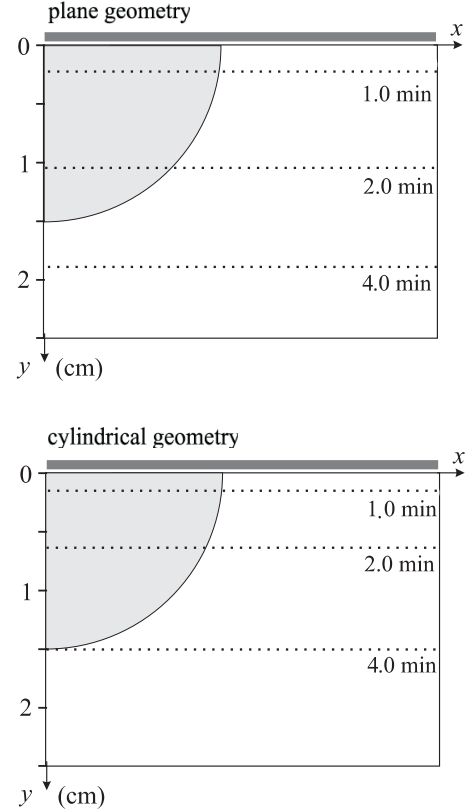}
\end{center}
\caption{The necrosis growth in living tissue without response, where the
existence of the tumor has no effect on the necrosis formation. The dotted
lines show the position of the necrosis interface at different moments of
time.} \label{Fig.3}
\end{figure}

\begin{figure}[hf]
\begin{center}
\includegraphics[width=60mm]{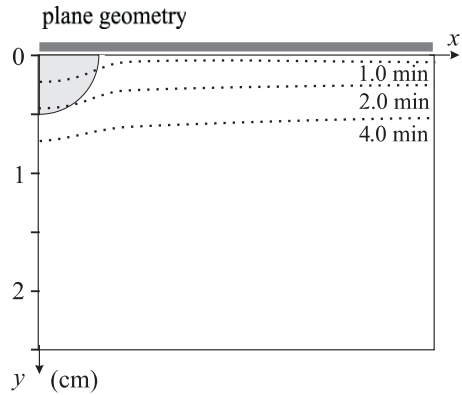}
\end{center}
\caption{The necrosis growth in living tissue containing a small tumor. The
tissue exhibits the strong response without delay. The dotted lines show the
position of the necrosis interface at different moments of time.} \label{Fig.4}
\end{figure}

In numerical simulation we have used the following typical values of the
thermal conductivity $\kappa \sim 7\cdot 10^{-3}$ \thinspace W/cm$\cdot $K,
the heat capacity $c\sim 3.5$\thinspace J/g$\cdot $ K, and the density $\rho
\sim 1$\thinspace g/cm$^{3}$ of the tissue, as well as set the blood
perfusion rate $j_{0}\sim 0.3$\thinspace min$^{-1}$ and the factors $f=0.5$,
$F=2$. Figure~\ref{Fig.3} demonstrates the necrosis growth in living tissue
without response. In this case the existence of the tumor has no effect on
the necrosis growth. A sufficiently small tumor also cannot substantially
affect the necrosis growth, as is seen in Fig.~\ref{Fig.4} exhibiting the
necrosis growth in living tissue with strong response without delay. The
tumor radius was chosen equal to $\mathcal{L}=0.5$~cm the tissue response to
heating was assumed to be strong, $\alpha =0.1$, and without delay, $\tau =0$%
. This case has been considered because under such conditions the effect of
large tumor on the necrosis growth is most pronounced.

\begin{figure}[fh]
\begin{center}
\includegraphics[width=60mm]{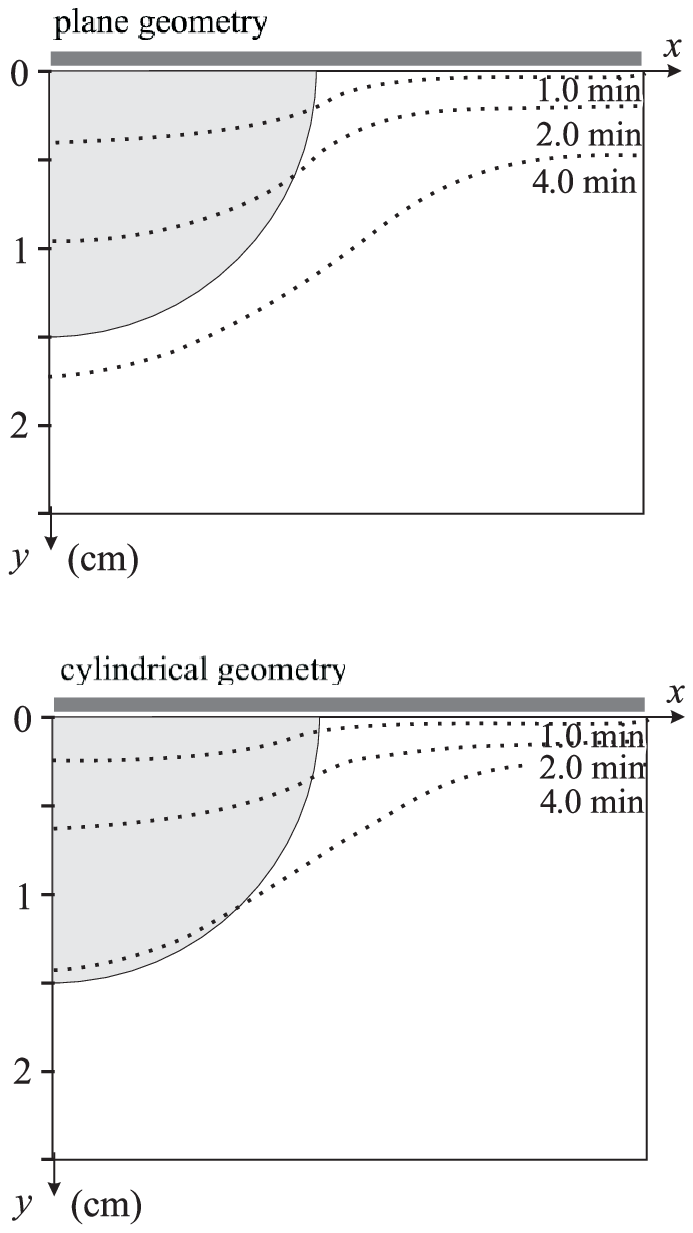}
\end{center}
\caption{The necrosis growth in living tissue containing a large tumor. The
tissue response to temperature increase is strong and not delayed. The dotted
lines show the position of the necrosis interface at different moments of
time.} \label{Fig.5}
\end{figure}

\begin{figure}[fh]
\begin{center}
\includegraphics[width=60mm]{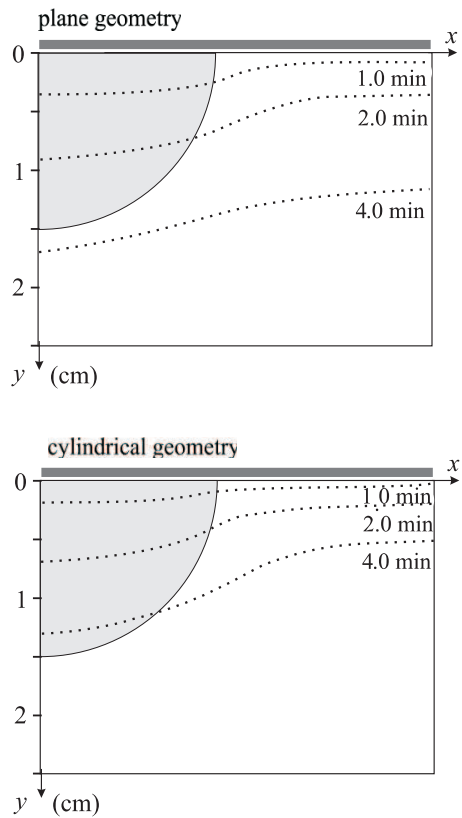}
\end{center}
\caption{The necrosis growth in living tissue containing a large tumor. The
tissue response to temperature increase is strong and delayed. The dotted
lines show the position of the necrosis interface at different moments of
time.} \label{Fig.6}
\end{figure}

When the tumor size is sufficiently large, $\mathcal{L}>\Re $, and the
tissue response is not too delayed, $\tau \ll \Re ^{2}/D$ the
self-localization phenomenon comes into being as demonstrated in Figures~\ref{Fig.5}--%
\ref{Fig.6}. Namely, Fig.~\ref{Fig.5} exhibits the necrosis growth in living
tissue containing a tumor of size of $\mathcal{L}=1.5$~cm where the tissue
response is strong, $\alpha =0.1$, and not delayed, $\tau =0$. Exactly this
situation matches the conditions under which the self-localization is most
pronounced. When the tissue response is delayed the effect of the tumor
existence is weakened as it is demonstrated in Fig.~\ref{Fig.6}, where we set
$\tau =1$~min.

\section{Closing remarks}

Concluding the paper we would like to state once more the obtained result.
Namely, we have analyzed the necrosis growth due to thermal coagulation
induced by laser light absorption and limited by heat diffusion into the
surrounding live tissue. The tissue is assumed to contain a certain tumor
where the blood perfusion rate does not change during the action. The latter
certainly concerns only the undamaged part of the tumor. By contrast, the
normal tissue responds strongly to increase in the tissue temperature and, as a
result, the blood perfusion rate can grow by tenfold.

We studied in detail the necrosis formation under conditions typical for a real
course of thermal therapy treatment, the duration of the action has been taken
about 5~minutes when the necrosis domain of size about or above 1~cm is
formed. In particular, we have shown that if the tumor size is sufficiently
large, it exceeds 1~cm, and the tissue response is not too delayed, the delay
time does not exceed~1 min, then there are conditions under which the relative
volume of the damaged normal tissue is small in comparison with the tumor
volume after the tumor is coagulated totally.

\begin{acknowledgments}
This work was supported by STCU grant \#1675.
\end{acknowledgments}

\end{document}